\documentclass[letter]{jpsj2}
\usepackage{mathptm,times}
\usepackage{amsmath,amssymb,amsthm}
\usepackage{latexsym}
\makeatletter

\newdimen\LENB \newdimen\LENW \newdimen\THI
\newdimen\LENWH \newdimen\LENTOT \newcount\N
\def\vbrknlnele#1#2#3{
  \LENB=#1pt \LENW=#2pt \THI=#3pt
  \LENWH=\LENW \divide\LENWH by 2
  \LENTOT=\LENB \advance\LENTOT by \LENW
  \vbox to \LENTOT{
    \vbox to \LENWH{}
    \nointerlineskip
    \vbox to \LENB{\hbox to \THI{\vrule width \THI height \LENB}}
    \nointerlineskip
    \vbox to \LENWH{}
  }}

\def\vbrknln#1{
  \N=#1
  \vcenter{
    \vbox{
      \loop\ifnum\N>0
        \vbox to 4pt{\vbrknlnele{2}{2}{0.1}}
        \nointerlineskip
        \advance\N by -1
      \repeat
  }}}

\def\hbrknlnele#1#2#3{
  \LENB=#1pt \LENW=#2pt \THI=#3pt
  \LENTOT=\LENB \advance\LENTOT by \LENW
  \vcenter{
    \vbox to \THI{
      \hbox to \LENTOT{
        \hfil
        \vrule width \LENB height \THI
        \hfil}
  }}}

\def\hblele{\hbrknlnele{2}{2.2}{0.1}}

\def\hblfil{\cleaders\hbox{$ \m@th \mkern1mu \hblele \mkern1mu $}\hfill} 
\makeatother

\title{
Construction of Integrals of Higher-Order Mappings
}
\author
{Ken-ichi \textsc{Maruno}$^{1,}$$^{2}$, 
and G. Reinout W. \textsc{Quispel}$^{3,}$$^{4}$
}
\inst{
$^{1}$Department of Mathematics, 
The University of Texas-Pan American, 
Edinburg, Texas 78539-2999, USA\\
$^{2}$Faculty of Mathematics, 
Kyushu University, Hakozaki, Higashi-ku, Fukuoka, 812-8581\\
$^{3}$
~Department of Mathematics, 
La Trobe University, Melbourne, Victoria 3086, Australia\\
$^{4}$
~Centre of Excellence for Mathematics and Statistics of 
Complex Systems, 
La Trobe University, Melbourne, Victoria 3086, Australia
}

\abst{
We find that certain higher-order mappings arise as reductions of 
the integrable discrete A-type KP (AKP) and B-type KP (BKP) equations. 
We find conservation laws
 for the AKP and BKP equations, then we use these conservation laws to derive
 integrals of the associated reduced maps. 
}

\kword{integrable mappings, integrals, bilinear equations}

\begin{document}
\maketitle




\def \part {\partial}
\def \ba {\begin{array}}
\def \ea {\end{array}}
\def \si {\sigma}
\def \al {\alpha}
\def \la {\lambda}

The search for discrete integrable systems has received considerable 
attention in the past decade. This has resulted in the discovery of 
integrable mappings of the second order, e.g., the
Quispel-Roberts-Thompson (QRT) mapping \cite{QRT}, 
and discrete Painlev\'e equations \cite{painleve}. Apart from 
second-order integrable mappings, 
the results for higher-order integrable mappings are few
\cite{HKY,Capel,Papa,QCPN,RQpreprint,Iatrou,TQT,FNC2005,Bellon}. 
Discrete integrable systems have applications to various areas of
physics, such as statistical mechanics, quantum gravity, and discrete
analogues of integrable systems in classical mechanics and solid-state
physics. 
Here, we study a novel class of higher-order integrable mappings that 
have bilinear forms. 

As an example, we discuss the following 6th-order mapping:
\begin{eqnarray}
&&Dx_{n+3}x_{n+2}^2x_{n+1}^3x_{n}^3x_{n-1}^3x_{n-2}^2x_{n-3}+
Ax_{n+2}x_{n+1}^2x_{n}^2x_{n-1}^2x_{n-2}\nonumber\\
&&\qquad +Bx_{n+1}x_nx_{n-1}+C=0\,.
\label{6mapBKP-1}
\end{eqnarray}
(Here and below $A, B, C, D$ are arbitrary parameters). 
How can we obtain integrals for this mapping? 
In ref. \citen{HKY}, a method for
the construction of integrals was proposed and integrable third-order
mappings that possess two integrals were obtained.  
However, this method is not applicable to higher-order mappings 
because this method uses some ansatz at first and requires the use of 
high-performance computers. If we consider 6th-order, 8th-order and 
higher-order mappings, this method does not work, because current computer
power is not sufficient \cite{Kimura}.  

In this Letter, we propose a systematic 
method of constructing integrals for a class
of higher-order integrable mappings that can be transformed into 
a single bilinear form without the use of computers. 
Our proposed method is based on discrete 
bilinear forms related to the A-type KP (AKP)
and B-type KP (BKP) soliton equations. Conservation laws for integrable 
partial difference equations have been studied in
refs. \citen{Hydon} and \citen{Hydon2}. 

Before we discuss conservation laws for discrete systems, let 
us briefly recall conservation laws for continuous systems
\cite{Olver}. 
To this end, consider a (scalar) partial differential equation (PDE) 
$\Delta[x,u^{(i)}]=0$. The conservation law of such a PDE is a divergence
expression 
\[
 \sum_{j}\frac{\partial P_j}{\partial x_j}=0\,,
\]
which vanishes for all solutions of the given system. It follows that 
there exists a function $\Lambda$ (called the {\it characteristic} of
the given conservation law) such that 
$\sum_{j}\frac{\partial P_j}{\partial x_j} = \Lambda \Delta$.
Similarly, the conservation law of a scalar partial difference equation 
$\Delta[n,u_n]=0$ is the expression
$\sum_j (S_j- {\it id})P_j=0$,
which vanishes for all solutions of the discrete system. 
(Here $S_j$ is a unit shift in the $n_j$ direction, and $\Delta[n,u_n]$
denotes a smooth function depending on $n$, $u_n$ and finitely many
iterates of $u_n$). 
Again there exists a function $\Lambda$ such that 
\begin{equation}
 \sum_j (S_j- {\it id})P_j=\Lambda \Delta\,.\label{characteristic}
\end{equation}
We will call $\Lambda$ the characteristic of the discrete 
conservation law. 

Here, we give a list of characteristics of the discrete AKP and BKP
equations. \\
{\bf Discrete BKP equation}\\
The discrete BKP (Miwa) equation \cite{Miwa} is given by 
\begin{eqnarray}
&&A\tau_{k+1,l,m}\tau_{k,l+1,m+1}
+B\tau_{k,l+1,m}\tau_{k+1,l,m+1}\nonumber\\
&&\, +C\tau_{k,l,m+1}\tau_{k+1,l+1,m}
+D\tau_{k,l,m}\tau_{k+1,l+1,m+1}=0\,.
\end{eqnarray}
We have found the following 6 explicit rational characteristics for 
the discrete BKP equation:
\begin{eqnarray*}
&&\Lambda_1=A\left(
\frac{f_1(k)\tau_{k-1,l+1,m+1}}{\tau_{k,l,m+1}\tau_{k,l+1,m+1}\tau_{k,l+1,m}} 
-
\frac{f_1(k+1)\tau_{k+2,l,m}}{\tau_{k+1,l,m}\tau_{k+1,l+1,m}\tau_{k+1,l,m+1}} 
\right) \\
&&\qquad +D\left(
\frac{f_1(k)\tau_{k-1,l,m}}{\tau_{k,l,m}\tau_{k,l+1,m}\tau_{k,l,m+1}} 
-
\frac{f_1(k+1)
\tau_{k+2,l+1,m+1}}{\tau_{k+1,l+1,m+1}\tau_{k+1,l+1,m}\tau_{k+1,l,m+1}} 
\right)\,,\\
  &&\Lambda_2=
B\left(
\frac{f_2(l)\tau_{k+1,l-1,m+1}}{\tau_{k+1,l,m}\tau_{k+1,l,m+1}\tau_{k,l,m+1}} 
-
\frac{f_2(l+1)\tau_{k,l+2,m}}{\tau_{k,l+1,m}\tau_{k,l+1,m+1}\tau_{k+1,l+1,m}} 
\right) \\
  &&\qquad +
D\left(
\frac{f_2(l)\tau_{k,l-1,m}}{\tau_{k,l,m}\tau_{k,l,m+1}\tau_{k+1,l,m}} 
-
\frac{f_2(l+1)
\tau_{k+1,l+2,m+1}}{\tau_{k+1,l+1,m+1}\tau_{k,l+1,m+1}\tau_{k+1,l+1,m}} 
\right)\,,\\
  &&\Lambda_3=
C\left(
\frac{f_3(m)\tau_{k+1,l+1,m-1}}{\tau_{k+1,l+1,m}\tau_{k,l+1,m}\tau_{k+1,l,m}} 
-
\frac{f_3(m+1)\tau_{k,l,m+2}}{\tau_{k,l,m+1}\tau_{k,l+1,m+1}\tau_{k+1,l,m+1}} 
\right) \\
  &&\qquad +
D\left(
\frac{f_3(m)\tau_{k,l,m-1}}{\tau_{k+1,l,m}\tau_{k,l+1,m}\tau_{k,l,m}} 
-
\frac{f_3(m+1)\tau_{k+1,l+1,m+2}}
{\tau_{k+1,l+1,m+1}\tau_{k,l+1,m+1}\tau_{k+1,l,m+1}} 
\right)\,,\\
  &&\Lambda_4=C\left(
\frac{f_4(l)\tau_{k,l-1,m+1}}{\tau_{k+1,l,m+1}\tau_{k,l,m+1}\tau_{k,l,m}} 
-
\frac{f_4(l+1)\tau_{k+1,l+2,m}}{\tau_{k+1,l+1,m}\tau_{k,l+1,m}\tau_{k+1,l+1,m+1}} 
\right) \\
  &&\qquad +A\left(
\frac{f_4(l)\tau_{k+1,l-1,m}}{\tau_{k+1,l,m}\tau_{k,l,m}\tau_{k+1,l,m+1}} 
-
\frac{f_4(l+1)\tau_{k,l+2,m+1}}
{\tau_{k,l+1,m+1}\tau_{k,l+1,m}\tau_{k+1,l+1,m+1}} 
\right)\,,\\
  &&\Lambda_5=A\left(
\frac{f_5(m)\tau_{k+1,l,m-1}}{\tau_{k+1,l+1,m}\tau_{k+1,l,m}\tau_{k,l,m}} 
-
\frac{f_5(m+1)\tau_{k,l+1,m+2}}{\tau_{k,l+1,m+1}\tau_{k,l,m+1}\tau_{k+1,l+1,m+1}} 
\right) \\
  &&\qquad +B\left(
\frac{f_5(m)\tau_{k,l+1,m-1}}{\tau_{k,l+1,m}\tau_{k,l,m}\tau_{k+1,l+1,m}} 
-
\frac{f_5(m+1)\tau_{k+1,l,m+2}}{\tau_{k+1,l,m+1}\tau_{k,l,m+1}\tau_{k+1,l+1,m+1}} 
\right)\,,\\
  &&\Lambda_6=B\left(
\frac{f_6(k)\tau_{k-1,l+1,m}}{\tau_{k,l+1,m+1}\tau_{k,l+1,m}\tau_{k,l,m}} 
-
\frac{f_6(k+1)\tau_{k+2,l,m+1}}{\tau_{k+1,l,m+1}\tau_{k+1,l,m}\tau_{k+1,l+1,m+1}} 
\right) \\
  &&\qquad +C\left(
\frac{f_6(k)\tau_{k-1,l,m+1}}{\tau_{k,l,m+1}\tau_{k,l,m}\tau_{k,l+1,m+1}} 
-
\frac{f_6(k+1)\tau_{k+2,l+1,m}}{\tau_{k+1,l+1,m}\tau_{k+1,l,m}\tau_{k+1,l+1,m+1}} 
\right)\,,
\end{eqnarray*}
where $f_1(x),\cdots,f_6(x)$ are arbitrary functions. 
(If we choose $f_j(x)$ as constants or $(-1)^x$, we can produce 12
characteristics. 
Throughout this Letter, we only consider reductions in which all independent
integrating factors are obtained taking $f_j(x)=${\rm
 constant}.)
From these characteristics, we can obtain the associated conservation
laws, using eq. (\ref{characteristic}). For example, $P_1, P_2$ and $P_3$ 
associated to $\Lambda_1$ are
\begin{eqnarray*}
  &&P_1=f_1(k)\Bigl[-A^2\frac{\tau_{k-1,l+1,m+1}\tau_{k+1,l,m}}
{\tau_{k+1,l+1,m}\tau_{k+1,l,m+1}}
-D^2\frac{\tau_{k-1,l,m}\tau_{k+1,l+1,m+1}}
{\tau_{k,l+1,m}\tau_{k,l,m+1}}
-AB\frac{\tau_{k-1,l+1,m}\tau_{k+1,l,m}}
{\tau_{k,l,m}\tau_{k,l+1,m}}
\\
  &&\quad -BD\frac{\tau_{k-1,l+1,m}\tau_{k+1,l+1,m+1}}
{\tau_{k,l+1,m+1}\tau_{k,l+1,m}}
-AC\frac{\tau_{k-1,l,m+1}\tau_{k+1,l,m}}
{\tau_{k,l,m}\tau_{k,l,m+1}}
-CD\frac{\tau_{k-1,l,m+1}\tau_{k+1,l+1,m+1}}
{\tau_{k,l+1,m+1}\tau_{k,l,m+1}}\\
  && \quad -AD\left(
\frac{\tau_{k-1,l,m}\tau_{k,l+1,m+1}\tau_{k+1,l,m}}
{\tau_{k,l,m}\tau_{k,l+1,m}\tau_{k,l,m+1}}
+\frac{\tau_{k,l,m}\tau_{k-1,l+1,m+1}\tau_{k+1,l+1,m+1}}
{\tau_{k,l+1,m+1}\tau_{k,l+1,m}\tau_{k,l,m+1}}
\right)\Bigr]\,,
\\
  &&P_2=f_1(k)\Bigl(
AC\frac{\tau_{k+1,l,m}\tau_{k-1,l,m+1}}
{\tau_{k,l,m+1}\tau_{k,l,m}}
-BD
\frac{\tau_{k-1,l,m}\tau_{k+1,l,m+1}}
{\tau_{k,l,m+1}\tau_{k,l,m}}\Bigr)\,,
\\
  &&P_3=f_1(k)\Bigl(
AB\frac{\tau_{k+1,l,m}\tau_{k-1,l+1,m}}
{\tau_{k,l,m}\tau_{k,l+1,m}}
-CD\frac{\tau_{k-1,l,m}\tau_{k+1,l+1,m}}
{\tau_{k,l+1,m}\tau_{k,l,m}}\Bigr)\,,
\end{eqnarray*}
where $P_1$, $P_2$ and $P_3$ correspond to the 
$k$-, $l$- and $m$-directions, respectively.
\quad \\
{\bf Discrete AKP equation}\\
The discrete AKP (Hirota-Miwa) equation \cite{Hirota,Miwa} is given by 
\begin{eqnarray}
&& A\tau_{k+1,l,m}\tau_{k,l+1,m+1}
+B\tau_{k,l+1,m}\tau_{k+1,l,m+1}\nonumber\\
&&\qquad +C\tau_{k,l,m+1}\tau_{k+1,l+1,m}=0\,.
\end{eqnarray}
Note that the discrete AKP equation is the special case of $D=0$ of the 
discrete BKP equation. 
The discrete AKP equation inherits the above 
6 characteristics (with $D=0$) from the discrete BKP equation, and 
we have found the following additional characteristic:
\begin{eqnarray*}
  &&\Lambda_7=
\frac{f_7(k+l+m)\tau_{k,l,m}}{\tau_{k+1,l,m}\tau_{k,l+1,m}\tau_{k,l,m+1}} 
-
\frac{f_7(k+l+m+1)\tau_{k+1,l+1,m+1}}{\tau_{k,l+1,m+1}
\tau_{k+1,l,m+1}\tau_{k+1,l+1,m}} 
\,, 
\end{eqnarray*}
where $f_7(x)$ is an arbitrary function.\\
{\bf Reduction to Finite-Dimensional 
Mappings and Construction of their Integrals}\\
{\bf First example:}
Consider the following 4th-order mapping: 
\begin{equation}
  Dx_{n+2}x_{n+1}^2x_{n}^2x_{n-1}^2x_{n-2}+
Ax_{n+1}x_nx_{n-1}+B+C/x_n=0\,.\label{4thBKP}
\end{equation}
Using the transformation of the dependent variable
$
x_n=\tau_{n+1}\tau_{n-1}/\tau_{n}^2\,, 
$
we obtain the bilinear form
\begin{equation}
D\tau_{n+3}\tau_{n-3}+A\tau_{n+2}\tau_{n-2}+B\tau_{n+1}\tau_{n-1}
+C\tau_n^2=0
\,.\label{4map-bi}
\end{equation}
This bilinear form is obtained from the discrete BKP equation by 
applying the reduction $\tau_{n}\equiv\tau_{Z_1k+Z_2l+Z_3m}$, where 
$Z_1=1, Z_2=2, Z_3=3$. Using the characteristics of the discrete BKP
equation, we obtain the following integrating factors for the discrete
bilinear form eq. (\ref{4map-bi}): 
\begin{eqnarray*}
  &&\Lambda_1=
-\Lambda_6\,,\quad 
\Lambda_2=
-\Lambda_4\,,\quad 
\Lambda_3=
\frac{A}{D}\Lambda_4+\frac{B}{D}\Lambda_6\,,\\
  &&\Lambda_4=C\left(
\frac{\tau_{n-2}}{\tau_{n+1}\tau_{n}\tau_{n-3}} 
-
\frac{\tau_{n+2}}{\tau_{n}\tau_{n-1}\tau_{n+3}} 
\right) +A\left(
\frac{\tau_{n-4}}{\tau_{n-2}\tau_{n-3}\tau_{n+1}} 
-
\frac{\tau_{n+4}}{\tau_{n+2}\tau_{n-1}\tau_{n+3}} 
\right)\,,\\
  &&\Lambda_5=
-\frac{A}{D}\Lambda_4-\frac{B}{D}\Lambda_6
\,,\\
  &&\Lambda_6=B\left(
\frac{\tau_{n-2}}{\tau_{n+2}\tau_{n-1}\tau_{n-3}} 
-
\frac{\tau_{n+2}}{\tau_{n+1}\tau_{n-2}\tau_{n+3}} 
\right) +C\left(
\frac{\tau_{n-1}}{\tau_{n}\tau_{n-3}\tau_{n+2}} 
-
\frac{\tau_{n+1}}{\tau_{n}\tau_{n-2}\tau_{n+3}} 
\right)\,.
\end{eqnarray*}
It is confirmed using the bilinear form eq. (\ref{4map-bi}) 
that the integrating factors $\Lambda_1$, $\Lambda_2$, $\Lambda_3$ 
and $\Lambda_5$
lead to the indicated linear combinations of $\Lambda_4$ and $\Lambda_6$. 
Note that the two integrating factors 
$\Lambda_4$ and $\Lambda_6$ are independent. 
From the above integrating factors, we can write the integrating factors 
in terms of the $x$-variable:
\begin{eqnarray*}
  &&\tilde{\Lambda}_4=\tau_{n-1}\tau_{n+1}\Lambda_4=
C\left(
\frac{1}{x_{n-1}x_{n-2}}-\frac{1}{x_{n+2}x_{n+1}}
\right) 
+A\left(x_{n-2}x_{n-3}-x_{n+3}x_{n+2}\right)\,,\\
  &&\tilde{\Lambda}_6=\tau_{n-1}\tau_{n+1}\Lambda_6\\
  && \quad =
B\left(
\frac{1}{x_{n+1}x_nx_{n-1}x_{n-2}}-\frac{1}{x_{n+2}x_{n+1}x_nx_{n-1}}
\right) 
+C\left(
\frac{1}{x_{n+1}x_nx_{n-1}^2x_{n-2}}-\frac{1}{x_{n+2}x_{n+1}^2x_nx_{n-1}}
\right)\,.
\end{eqnarray*}
We then obtain the following two integrals:
\begin{eqnarray*}
  &&Q_4=CDx_{n+2}x_{n+1}^2x_n^2x_{n-1}
-ADx_{n+3}x_{n+2}^2x_{n+1}^2x_n^2x_{n-1}^2x_{n-2}\\
  &&\quad
 -A^2(x_{n+3}x_{n+2}x_{n+1}x_nx_{n-1}+x_{n+2}x_{n+1}x_nx_{n-1}x_{n-2})
-C^2\left(\frac{1}{x_{n+2}x_{n+1}x_n}
+\frac{1}{x_{n+1}x_nx_{n-1}}\right)
\\
  &&\quad 
-BC\left(\frac{1}{x_{n+2}x_{n+1}}
+\frac{1}{x_{n+1}x_n}+\frac{1}{x_nx_{n-1}}\right)
-AB\sum_{j=0}^4x_{n+3-j}x_{n+2-j}-AC\sum_{j=0}^3
\frac{x_{n+3-j}x_{n+2-j}}{x_{n-j}}\,,\\
  &&Q_6=BDx_{n+2}x_{n+1}x_nx_{n-1}+CD(x_{n+2}x_{n+1}x_n+x_{n+1}x_nx_{n-1})
-AB\left(\frac{1}{x_{n+2}}+\frac{1}{x_{n+1}}
+\frac{1}{x_n}+\frac{1}{x_{n-1}}\right)\\
  &&\quad -AC\left(\frac{1}{x_{n+2}x_{n+1}}
+\frac{1}{x_{n+1}x_n}+\frac{1}{x_nx_{n-1}}\right)
-\frac{B^2}{x_{n+2}x_{n+1}x_nx_{n-1}}
-\frac{C^2}{x_{n+2}x_{n+1}^2x_n^2x_{n-1}}\\
  &&\quad -BC\left(\frac{1}{x_{n+2}x_{n+1}x_n^2x_{n-1}}
+\frac{1}{x_{n+2}x_{n+1}^2x_nx_{n-1}}\right)\,.
\end{eqnarray*}
It is not difficult to show that $Q_4$ and $Q_6$ are functionally
independent. 

In the special case of $D=0$, the fourth-order mapping 
eq. (\ref{4thBKP}) reduces to the 
second-order mapping
\begin{equation}
Ax_{n+1}x_nx_{n-1}+B+C/x_n=0\,,\label{2nd-AKP}
\end{equation}
which is a special case of the QRT mapping \cite{QRT}. 
Using the transformation of the dependent variable
$
x_n=\tau_{n+1}\tau_{n-1}/\tau_{n}^2\,, 
$
we obtain the bilinear form
\begin{equation}
A\tau_{n+2}\tau_{n-2}+B\tau_{n+1}\tau_{n-1}
+C\tau_n^2=0
\,.\label{2map-bi}
\end{equation}
This bilinear form is obtained from the discrete AKP equation by 
applying the reduction $\tau_{n}\equiv\tau_{Z_1k+Z_2l+Z_3m}$, where 
$Z_1=1, Z_2=2, Z_3=3$. Using the characteristics of the discrete AKP
equation, we obtain the following integrating factors for the discrete
bilinear form eq. (\ref{2map-bi}):
\begin{eqnarray*}
  &&\Lambda_1=\left(
\frac{\tau_{n+1}}{\tau_{n}\tau_{n+2}\tau_{n-1}} 
-
\frac{\tau_{n-1}}{\tau_{n-2}\tau_{n}\tau_{n+1}} 
\right)\,,\quad 
\Lambda_2=
-\Lambda_1\,,\\
&&\Lambda_3=
\frac{C}{A}\Lambda_1\,,\quad
\Lambda_4=
B\Lambda_1\,,\quad
\Lambda_5=
-\frac{C^2}{A}\Lambda_1\,,\\
  &&\Lambda_6=
-A\Lambda_1\,,\quad
\Lambda_7=
\frac{C}{A}\Lambda_1\,.
\end{eqnarray*} 
There is only one independent integrating factor, $\Lambda_1$. 
From $\Lambda_1$, we can write the integrating factor 
in terms of the $x$-variable:
$\tilde{\Lambda}_1=\tau_{n+1}\tau_{n-1}\Lambda_1=
1/x_{n+1}-1/x_{n-1}$.
We then obtain the following integral:
\begin{eqnarray*}
Q_1=-Ax_{n+1}x_n+B\left(\frac{1}{x_{n+1}}+\frac{1}{x_n}\right)
+\frac{C}{x_{n+1}x_n}\,.
\end{eqnarray*}

{\bf Second example:}
As a second example, 
let us discuss the following 6th-order mapping:
\begin{equation}
  Dx_{n+3}x_{n+2}^2x_{n+1}^3x_{n}^3x_{n-1}^3x_{n-2}^2x_{n-3}+
Ax_{n+2}x_{n+1}^2x_{n}^2x_{n-1}^2x_{n-2}+Bx_{n+1}x_nx_{n-1}+C=0\,.
\label{6mapBKP}
\end{equation}
Using the transformation of the dependent variable
$
x_n=\tau_{n+1}\tau_{n-1}/\tau_{n}^2\,, 
$
we obtain the bilinear form
\begin{equation}
A\tau_{n+3}\tau_{n-3}+B\tau_{n+2}\tau_{n-2}
+C\tau_{n+1}\tau_{n-1}+D\tau_{n+4}\tau_{n-4}=0\,.\label{6mapBKP-bi}
\end{equation}
This bilinear form is obtained from the discrete BKP equation by 
applying the reduction $\tau_{n}\equiv\tau_{Z_1k+Z_2l+Z_3m}$ where 
$Z_1=1, Z_2=2, Z_3=5$ or $Z_1=1, Z_2=3, Z_3=4$. 
Using the characteristics of the discrete BKP
equation, we obtain the following integrating factors for the discrete
bilinear form eq. (\ref{6mapBKP-bi}): 
\begin{eqnarray*}
  &&\Lambda_1=
-\Lambda_6\,,\quad
\Lambda_2=
\Lambda_4\,,\quad
\Lambda_3=
-\Lambda_5\,,\\
  &&\Lambda_4=C\left(
\frac{\tau_{n-1}}{\tau_{n+2}\tau_{n+1}\tau_{n-4}} 
-
\frac{\tau_{n+1}}{\tau_{n-1}\tau_{n-2}\tau_{n+4}}\right)
+A\left(
\frac{\tau_{n-5}}{\tau_{n-3}\tau_{n-4}\tau_{n+2}} 
-
\frac{\tau_{n+5}}{\tau_{n+3}\tau_{n-2}\tau_{n+4}} 
\right)\,,\\
  &&\Lambda_5=A\left(
\frac{\tau_{n-8}}{\tau_{n-1}\tau_{n-3}\tau_{n-4}} 
-
\frac{\tau_{n+8}}{\tau_{n+3}\tau_{n+1}\tau_{n+4}} 
\right)
+B\left(
\frac{\tau_{n-7}}{\tau_{n-2}\tau_{n-4}\tau_{n-1}} 
-
\frac{\tau_{n+7}}{\tau_{n+2}\tau_{n+1}\tau_{n+4}} 
\right)\,,\\
  &&\Lambda_6=B\left(
\frac{\tau_{n-3}}{\tau_{n+3}\tau_{n-2}\tau_{n-4}} 
-
\frac{\tau_{n+3}}{\tau_{n+2}\tau_{n-3}\tau_{n+4}} 
\right)+C\left(
\frac{\tau_{n}}{\tau_{n+1}\tau_{n-4}\tau_{n+3}} 
-
\frac{\tau_{n}}{\tau_{n-1}\tau_{n-3}\tau_{n+4}} 
\right)\,.
\end{eqnarray*}
It is confirmed using the bilinear form eq. (\ref{6mapBKP-bi}) 
that integrating factors $\Lambda_1$, $\Lambda_2$ and $\Lambda_3$ 
lead to $\Lambda_6$, $\Lambda_4$ and $\Lambda_5$ respectively. 
Note that the 3 integrating factors 
$\Lambda_4$, $\Lambda_5$ and $\Lambda_6$ are independent. 
From the above integrating factors, we can write the integrating factors 
in terms of the $x$-variable:
\begin{eqnarray*}
  &&\tilde{\Lambda}_4=\tau_{n+1}\tau_{n-1}\Lambda_4\\
  && \quad =\frac{C}{x_{n+1}x_n^2x_{n-1}}
\left(
\frac{1}{x_{n-1}^2x_{n-2}^2x_{n-3}}-
\frac{1}{x_{n+3}x_{n+2}^2x_{n+1}^2}
\right)+\frac{A}{x_{n+1}x_nx_{n-1}}
\left(x_{n-3}x_{n-4}-x_{n+4}x_{n+3}\right)\,,\\
  &&\tilde{\Lambda}_5=\tau_{n+1}\tau_{n-1}\Lambda_5
=
Ax_{n}(x_{n-1}^2x_{n-2}^3x_{n-3}^4x_{n-4}^4x_{n-5}^3x_{n-6}^2x_{n-7}
-x_{n+7}x_{n+6}^2x_{n+5}^3x_{n+4}^4x_{n+3}^4x_{n+2}^3x_{n+1}^2
)\\
  &&\qquad \qquad 
+Bx_n(x_{n-1}^2x_{n-2}^3x_{n-3}^3x_{n-4}^3x_{n-5}^2x_{n-6}
-x_{n+6}x_{n+5}^2x_{n+4}^3x_{n+3}^3x_{n+2}^3x_{n+1}^2)
\,,\\
  &&\tilde{\Lambda}_6=\tau_{n+1}\tau_{n-1}\Lambda_6
=\frac{B}{x_{n+2}x_{n+1}^2x_n^2x_{n-1}^2x_{n-2}}
\left(
\frac{1}{x_{n-3}}-\frac{1}{x_{n+3}}
\right)\\
  && \qquad \qquad +\frac{C}
{x_{n+2}x_{n+1}^2x_n^3x_{n-1}^2x_{n-2}}
\left(
\frac{1}{x_{n-1}x_{n-2}x_{n-3}}
-\frac{1}{x_{n+3}x_{n+2}x_{n+1}}
\right)
\,.
\end{eqnarray*}
We then obtain the following three integrals. (Note that eq. 
(\ref{6mapBKP}) can be used to eliminate, e.g., $x_{n+4}$ and $x_{n-3}$
from $Q_4$, and similarly for $Q_5$ and $Q_6$.):
\begin{eqnarray*}
 &&Q_4=CD(x_{n+3}x_{n+2}^2x_{n+1}^2x_n+x_{n+2}x_{n+1}^2x_{n}^2x_{n-1}
+x_{n+1}x_{n}^2x_{n-1}^2x_{n-2})\\
  &&\quad 
-ADx_{n+4}x_{n+3}^2x_{n+2}^2x_{n+1}^2x_n^2x_{n-1}^2x_{n-2}^2x_{n-3}
-A^2x_{n+3}x_{n+2}x_{n+1}x_{n}x_{n-1}x_{n-2}(x_{n+4}+x_{n-3})\\
  && \quad -BC\sum_{j=0}^{2}
\frac{1}{x_{n+3-j}x_{n+2-j}^2x_{n+1-j}^2x_{n-j}}
-AB\sum_{j=0}^{6}
x_{n+4-j}x_{n+3-j}
\\
  && \quad -C^2\left(
\frac{1}{x_{n+3}x_{n+2}^2x_{n+1}^3x_{n}^2x_{n-1}}
+\frac{1}{x_{n+2}x_{n+1}^2x_{n}^3x_{n-1}^2x_{n-2}}
\right)-AC
\sum_{j=0}^{3}
\frac{x_{n+4-j}x_{n+3-j}}{x_{n+1-j}x_{n-j}x_{n-1-j}}\,,
\\
  &&Q_5=
AD\sum_{j=0}^{3}x_{n+7-j}x_{n+6-j}^2x_{n+5-j}^3x_{n+4-j}^4
x_{n+3-j}^5x_{n+2-j}^5
x_{n+1-j}^5x_{n-j}^4x_{n-1-j}^3x_{n-2-j}^2x_{n-3-j}\\
  && \quad +BD\sum_{j=0}^{2}x_{n+6-j}x_{n+5-j}^2x_{n+4-j}^3
x_{n+3-j}^4x_{n+2-j}^4
x_{n+1-j}^5x_{n-j}^4x_{n-1-j}^3x_{n-2-j}^2x_{n-3-j}\\
  && \quad +A^2\sum_{j=0}^{4}x_{n+7-j}x_{n+6-j}^2x_{n+5-j}^3x_{n+4-j}^4
x_{n+3-j}^4x_{n+2-j}^4
x_{n+1-j}^4x_{n-j}^3x_{n-1-j}^2x_{n-2-j}\\
  && \quad +AB\sum_{j=0}^{4}
x_{n+6-j}x_{n+5-j}^2x_{n+4-j}^3
x_{n+3-j}^3x_{n+2-j}^4
x_{n+1-j}^4x_{n-j}^3x_{n-1-j}^2x_{n-2-j}\\
  &&\quad +AB
\sum_{j=0}^{4}x_{n+7-j}x_{n+6-j}^2x_{n+5-j}^3x_{n+4-j}^4
x_{n+3-j}^4x_{n+2-j}^3
x_{n+1-j}^3x_{n-j}^2x_{n-1-j}\\
  && \quad +B^2
\sum_{j=0}^{5}
x_{n+6-j}x_{n+5-j}^2x_{n+4-j}^3
x_{n+3-j}^3x_{n+2-j}^3
x_{n+1-j}^3x_{n-j}^2x_{n-1-j}\\
  &&\quad +BC\sum_{j=0}^{5}
x_{n+6-j}x_{n+5-j}^2x_{n+4-j}^3
x_{n+3-j}^3x_{n+2-j}^3
x_{n+1-j}^2x_{n-j}\\
  &&\quad  +AC 
\sum_{j=0}^{6}x_{n+7-j}x_{n+6-j}^2x_{n+5-j}^3x_{n+4-j}^4
x_{n+3-j}^4x_{n+2-j}^3
x_{n+1-j}^2x_{n-j}\,,
\\
  &&Q_6=
BDx_{n+3}x_{n+2}x_{n+1}x_{n}x_{n-1}x_{n-2}
+CD\sum_{j=0}^3x_{n+3-j}x_{n+2-j}x_{n+1-j}\\
  && \quad -AB\sum_{j=0}^5\frac{1}{x_{n+3-j}}
-AC\sum_{j=0}^2\frac{1}{x_{n+3-j}x_{n+2-j}x_{n+1-j}x_{n-j}}\\
  && \quad -\frac{B^2}{x_{n+3}x_{n+2}x_{n+1}x_{n}x_{n-1}x_{n-2}}
-\frac{C^2}{x_{n+3}x_{n+2}^2x_{n+1}^3x_{n}^3x_{n-1}^2x_{n-2}}\\
  && \quad -\frac{BC}{x_{n+3}x_{n+2}x_{n+1}^2x_{n}^2x_{n-1}^2x_{n-2}}
-\frac{BC}{x_{n+3}x_{n+2}^2x_{n+1}^2x_{n}^2x_{n-1}x_{n-2}}\,.
\end{eqnarray*}
Using a Mathematical software package such 
as Mathematica, one can show that $Q_4$, $Q_5$ and $Q_6$ are
functionally independent.

We consider the 4th-order mapping
\begin{equation}
Ax_{n+2}x_{n+1}^2x_{n}^2x_{n-1}^2x_{n-2}
+Bx_{n+1}x_nx_{n-1}+C=0\,.\label{4mapAKP}
\end{equation}
This mapping is the special case of $D=0$ of the 6th-order mapping
eq. (\ref{6mapBKP}).
Applying
$
x_n=\tau_{n+1}\tau_{n-1}/\tau_{n}^2\,, 
$
we obtain the bilinear form
\begin{equation}
A\tau_{n+3}\tau_{n-3}+B\tau_{n+2}\tau_{n-2}+C\tau_{n+1}\tau_{n-1}=0\,. 
\label{4mapAKP-bi}
\end{equation}
This bilinear form is obtained from the discrete AKP equation by 
applying the reduction $\tau_{n}\equiv\tau_{Z_1k+Z_2l+Z_3m}$, where 
$Z_1=1, Z_2=2, Z_3=5$ or $Z_1=1, Z_2=3, Z_3=4$. 
Using the characteristics of the discrete AKP
equation, we obtain the following integrating factors for the discrete
bilinear form eq. (\ref{4mapAKP-bi}): 
\begin{eqnarray*}
  &&\Lambda_1=
\frac{\tau_{n+2}}{\tau_{n+1}\tau_{n+3}\tau_{n-2}} 
-
\frac{\tau_{n-2}}{\tau_{n-3}\tau_{n-1}\tau_{n+2}} 
\,,\\
  &&\Lambda_2=
\frac{\tau_{n}}{\tau_{n-3}\tau_{n+2}\tau_{n+1}} 
-
\frac{\tau_{n}}{\tau_{n-2}\tau_{n+3}\tau_{n-1}} 
\,,\\
  &&\Lambda_3=
-\frac{C^2}{A^2}\Lambda_1-\frac{B^2C}{A^3}\Lambda_2\,,
\quad \Lambda_4=
-B\Lambda_2\,,\\
  &&\Lambda_5=
\frac{C^3}{A^2}\Lambda_1+\frac{B^2C^2}{A^3}\Lambda_2\,,
\quad \Lambda_6=
-A\Lambda_1\,,\quad \Lambda_7=
-\frac{C}{A}\Lambda_2\,. 
\end{eqnarray*}
It is confirmed using the bilinear form eq. (\ref{4mapAKP-bi}) 
that the two integrating factors $\Lambda_1$ and $\Lambda_2$ 
are independent. 
From the above integrating factors, we can write the integrating factors 
in terms of the $x$-variable:
\begin{eqnarray*}
&&\tilde{\Lambda}_1=\tau_{n+1}\tau_{n-1}\Lambda_1
=\frac{1}{x_{n+1}x_{n}x_{n-1}}
\left(\frac{1}{x_{n+2}} 
-\frac{1}{x_{n-2}} 
\right)\,,\\
&&\tilde{\Lambda}_2=\tau_{n+1}\tau_{n-1}\Lambda_2
=\frac{1}{x_{n+1}x_{n}^2x_{n-1}}
\left(\frac{1}{x_{n-1}x_{n-2}} 
-\frac{1}{x_{n+2}x_{n+1}} 
\right)\,.
\end{eqnarray*}
We then obtain the following two integrals:
\begin{eqnarray*}
  &&Q_1=
-Ax_{n+2}x_{n+1}x_nx_{n-1}+B\left(\frac{1}{x_{n+2}}+
\frac{1}{x_{n+1}}+\frac{1}{x_n}+\frac{1}{x_{n-1}}\right)
+\frac{C}{x_{n+2}x_{n+1}x_nx_{n-1}} \,,\\
  &&Q_2=
A(x_{n+2}x_{n+1}+x_{n+1}x_{n}+x_nx_{n-1})
-B\left(\frac{1}{x_{n+2}x_{n+1}x_n}+\frac{1}{x_{n+1}x_nx_{n-1}}\right)
-\frac{C}{x_{n+2}x_{n+1}^2x_{n}^2x_{n-1}}\,.
\end{eqnarray*}
We note that eq. (\ref{4mapAKP}) preserves the symplectic structure 
\[
\left(
\begin{array}{cccc}
0 & x_{n-2}x_{n-1} & -x_{n-2}x_n & x_{n-2}x_{n+1}\cr
-x_{n-1}x_{n-2} & 0 & x_{n-1}x_n & -x_{n-1}x_{n+1}\cr
x_n x_{n-2} & -x_nx_{n-1} & 0 & x_n x_{n+1} \cr
-x_{n+1}x_{n-2} & x_{n+1}x_{n-1} & -x_{n+1}x_n &0 \cr
\end{array}
\right)\,, 
\]
and that the two integrals $Q_1$ and $Q_2$ are in involution w.r.t. this
structure, giving an independent confirmation of the integrability of
eq. (\ref{4mapAKP}). 

We have studied a class of integrable mappings that can be 
transformed into a single bilinear form. 
We have proposed a method of constructing integrals of these 
higher-order integrable maps. The key to the construction is the 
conservation laws of the discrete bilinear forms of the associated 
AKP and BKP equations. 
Note that, generalizing the examples in this Letter, 
we can construct a family of higher-order
mappings from the discrete AKP and BKP equations, by 
applying the reduction $\tau_{n}\equiv\tau_{Z_1k+Z_2l+Z_3m}$ for any
$Z_1$, $Z_2$ and $Z_3$. 

In general, soliton equations in the 
1+1-dimension have infinitely many conservation laws. 
One may expect the existence of infinitely many conservation laws 
for discrete AKP and BKP equations.  
However, finding infinitely many conservation laws in the form of
eq. (2) 
is difficult in the case of 2+1-dimensional 
discrete soliton equations. Finding infinitely many conservation laws 
for discrete AKP and BKP equations is still an open problem.

We hope to discuss details 
of our methods and higher-order mappings in the class given
here in a forthcoming paper. 

\section*{Acknowledgments}

We acknowledge helpful discussions with K. Kimura. 
We gratefully acknowledge financial support from 
the Australian Research Council Centre of Excellence for 
Mathematics and Statistics of Complex Systems. 
KM also acknowledges support from the 21st Century 
COE program ``Development of Dynamic Mathematics with High Functionality'' 
at the Faculty of Mathematics, Kyushu University. 
KM thanks the Department of Mathematics, La Trobe University 
for thier warm hospitality.

\end{document}